\documentclass[superscriptaddress,aps,prd,twocolumn]{revtex4-1}
\usepackage[english]{babel}
\usepackage[utf8]{inputenc}
\usepackage{float}
\usepackage{amsmath}
\usepackage{amssymb}
\usepackage{graphicx}
\usepackage{color}

\usepackage{hyperref}

\usepackage[normalem]{ulem}

\begin{document}
	
\title{Constraints on high density equation of state from maximum neutron star mass}

\author{Márcio Ferreira}
\email{marcio.ferreira@uc.pt}
\affiliation{CFisUC, 
	Department of Physics, University of Coimbra, P-3004 - 516  Coimbra, Portugal}

\author{Constança Providência}
\email{cp@uc.pt}
\affiliation{CFisUC, 
	Department of Physics, University of Coimbra, P-3004 - 516  Coimbra, Portugal}

\date{\today}

\begin{abstract}
The low density nuclear matter equation of state is strongly constrained by nuclear properties, however, for constraining  the high density equation of state it is necessary to resort to indirect information obtained from the observation of neutron stars, compact objects that may have a central density several times nuclear matter saturation density, $n_0$. Taking a meta-modelling approach to generate a huge  set of equation of state that satisfy nuclear matter properties close to $n_0$ and that do not contain a first order phase transition, the possibility of constraining the high density equation of state was investigated. The entire information obtained from the GW170817 event for the probability distribution of $\tilde{\Lambda}$ was used to make a probabilistic inference of the EOS, which goes beyond the constraints imposed by nuclear matter properties.
Nuclear matter properties close to saturation, below $2n_0$, do not allow us to distinguish between equations of state that predict different neutron star (NS) maximum masses. This is, however, not true if the equation of state is constrained at low densities by the tidal deformability of the  NS merger associated to GW170817. Above $3n_0$, differences may be large, for both approaches,  and, in particular, the pressure and speed of sound of the  sets studied  do not overlap, showing that the knowledge of the NS maximum mass may give important information on the high density EOS.  Narrowing the maximum mass uncertainty interval will have a sizeable effect on constraining the  high density EOS.
\end{abstract}

\maketitle

\section{Introduction}

Neutron stars (NS) are special astrophysical objects
through which the properties of cold super-dense neutron-rich nuclear
matter may be investigated. The massive NSs observed, e.g., PSR J1614$-$2230 with $M=1.908\pm0.016\, M_\odot$ \cite{Arzoumanian2017,Fonseca2017,Demorest2010}, have
established strong constraints on the equation of state (EOS) of
nuclear matter. Further pulsars' observations, 
PSR J0348$+$0432 with $M=2.01 \pm 0.04 M_\odot$
\cite{Antoniadis2013} and MSP J0740$+$6620,
with a mass 
$2.08^{+0.07}_{-0.07}M_\odot$ 
\cite{Cromartie2019,Fonseca2021}, and  radius $12.39^{+1.30}_{-0.98}$ km obtained from NICER and XMM-Newton data  \cite{Riley2021} (together with an updated mass $2.072 ^{+0.067}_{-0.066} M_\odot$ ),
have strengthened the already  stiff constraints on the EOS. 
The NICER mission \cite{NICER} has estimated the mass and radius of the pulsar  PSR
J0030-0451:   respectively, ${1.34}_{-0.16}^{+0.15}\,{M}_{\odot }$, ${12.71}_{-1.19}^{+1.14}\,\mathrm{km}$ \cite{Riley19}, and ${1.44}_{-0.15}^{+0.15}\,{M}_{\odot }$, ${13.02}_{-1.06}^{+1.24}\,\mathrm{km}$ \cite{Miller19}.

Gravitational waves (GWs) are another crucial source of information about NS matter. 
GWs are emitted during the coalescence of binary 
NS systems and carry important information on the high density 
properties of the EOS. 
The analysis of the event GW170817 has settled
an upper bound on the effective tidal deformability of the binary $\tilde{\Lambda}$
\cite{TheLIGOScientific:2017qsa}. 
Using a low-spin prior, which is consistent
with the observed NS population, the value $\tilde{\Lambda}\le 800$ (90\% confidence) was determined from the GW170817 event. 
Tighter constraints were found in a follow up reanalysis \cite{Abbott:2018wiz}, 
with $\tilde{\Lambda}=300^{+420}_{-230}$ (90\% confidence), under minimal assumptions about the nature of the compact objects.
The two NS radii for the GW170817 event were estimated in \cite{Abbott18}, 
under the hypothesis that both NS are described by the same EOS and have spins within the range observed in Galactic binary NSs, to be $R_1=11.9^{+1.4}_{-1.4}$ km (heavier star) and $R_2=11.9_{-1.4}^{+1.4}$ km (lighter star). These constraints on $R_{1,2}$ were obtained requiring that the EOS supports NS with masses larger than $1.97M_{\odot}$.

The detection of GWs from the GW170817 event was followed by electromagnetic
counterparts,  the gamma-ray burst (GRB) GRB170817A \cite{grb}, and the
electromagnetic transient AT2017gfo \cite{kilo}, that set extra
constraints on the lower limit of the tidal deformability
\cite{Radice2017,Radice2018,Bauswein2019,Coughlin2018,Wang2018}. This last
constraint  seems to rule out very soft
EOS: the lower limit of  the tidal deformability of  a 1.37$M_\odot$
star set by the above studies limits the  tidal deformability
to  $\Lambda_{1.37M_\odot} > 210$ \cite{Bauswein2019}, 300 \cite{Radice2018},
279 \cite{Coughlin2018}, and 309 \cite{Wang2018}.
  
The LIGO/Virgo collaboration has recently reported 
the gravitational-wave observation of a compact binary coalescence (GW190814) \cite{Abbott:2020khf}. While the primary component of GW190814 is conclusively a black hole of mass $22.2-24.3M_{\odot}$, the secondary component of the binary with a mass  $2.50-2.67M_{\odot}$ remains yet inconclusive, which might be either the heaviest neutron star or the lightest black hole ever discovered in a double compact-object system \cite{Abbott:2020khf}. The absence of measurable tidal deformations and the absence of an electromagnetic counterpart from the GW190814 event are consistent both scenarios \cite{Abbott:2020khf}. The uncertainty on the nature of the second component have prompted an ongoing debate on whether the EOS of nuclear matter is able to accommodate such a massive NS.\\

In the present study, we will analyse which information on the high density EOS can be drawn from the knowledge of the NS maximum mass.  Several works have estimated NS maximum mass using different approaches taking into account the GW170817 observation,  the electromagnetic follow-up or/and  NICER  PSR J0030+0451 data  \cite{Shibata2017,Margalit2017,Alsing2018,Rezzolla2017,Shibata2019,AngLi2021}: the upper limit was fixed at $\approx 2.15-2.25M_\odot$ in \cite{Shibata2017}, and updated to $M_{\text{max}}<2.3 M_\odot$ in \cite{Shibata2019};  by the interval $2.01^{+0.04}_{-0.04}\le M_{\text{max}} / M_\odot \le 2.26^{+0.17}_{-0.15}$ was obtained for a nonrotating NS in \cite{Rezzolla2017}; \cite{Margalit2017} places the upper limit ${M}_{\max }\lesssim 2.17\,{M}_{\odot }$ (90\%); in \cite{Alsing2018} the limits $2.0 <M_{\text{max}} / M_\odot<2.2$ (68\%) and $2.0 <M_{\text{max}} / M_\odot<2.6 M_\odot$ (90\%) have been obtained; in \cite{Ruiz2017} the maximum mass $2.16 - 2.28M_\odot$ was calculated; in a recent study considering hybrid stars and both the GW170817 and the NICER observations the limits $2.36_{-0.26}^{+0.49}M_\odot$ and $2.39_{-0.28}^{+0.47}M_\odot$, taking two different hadronic models,  where obtained at 90\% credible interval  \cite{AngLi2021}. Considering that a black-hole was formed in GW170817 and taking conservative assumptions the authors of \cite{Annala:2021gom} concluded that the maximum mass should be $<2.53M_\odot$.

We assume that no first order phase transition occurs. Although restrictive, this hypothesis is supported by recent studies.  In \cite{Legred2021}, a one branch EOS was favored considering  a non-parametric EOS model based on Gaussian processes and taking as constraints  radio, x-ray and GW NS data, including the mass and radius measurement of the pulsar J0740$+$6620. A similar result was obtained in \cite{Pang2021}, where a first-order phase transition inside NSs was shown to be not favored, although not ruled out.

Nuclear matter properties are reasonably well constrained for densities below twice the saturation density by nuclear experiments. Therefore, as a second assumption, we take this information into account by considering a set of meta-EOS that satisfy experimental nuclear matter constraints below $\approx 2n_0$ and give rise to thermodynamically consistent EOSs that describe 1.97 $M_\odot$ NSs. 
Applying a probabilistic approach, we analyze how the information contained in the entire posterior probability distribution (pdf) for the binary tidal deformability of the GW170817 event \cite{Abbott:2018exr} affects the EOS pressure and speed of sound at high densities. It will be shown that non-overlapping 90\% credible intervals are obtained when different maximum masses are considered, indicating that a maximum mass constraint gives information on the star EOS. 
However, we should keep in mind that there is always the possibility that the ''real'' nuclear matter EOS consists in an unlikely realization for any probabilistic analysis approach, and thus the  ''real'' EOS might actually be realized in a low probability region. It is the nuclear force that determines the characteristics of the EOS. Having this in mind in most of our plots we represent the extremes of our sets.\\

The paper is organized as follows. The EOS parametrization and the method for generating the EOSs are presented in Sec. \ref{sec:EOS}. The results are discussed in Sec. \ref{sec:results}, where we analyze the properties of the different EOS data sets and perform a probabilistic inference on the high density region of the EOS. 
Finally, the conclusions are drawn in Sec. \ref{sec:conclusions}.

\section{EOS modelling}
\label{sec:EOS}
We assume the generic functional form for the energy per particle of homogeneous nuclear matter 
\begin{equation}
{\cal E}(n,\delta)=e_{\text{sat}}(n)+e_{\text{sym}}(n)\delta^2
\end{equation}
with 
\begin{align}
e_{\text{sat}}(n)&=E_{\text{sat}}+\frac{1}{2}K_{\text{sat}}x^2+\frac{1}{6}Q_{\text{sat}}x^3+\frac{1}{24}Z_{\text{sat}}x^4\\
e_{\text{sym}}(n)&=E_{\text{sym}}+L_{\text{sym}}x+\frac{1}{2}K_{\text{sym}}x^2+\frac{1}{6}Q_{\text{sym}}x^3\nonumber\\
&+\frac{1}{24}Z_{\text{sym}}x^4
\end{align}
where $x=(n-n_{0})/(3n_{0})$. 
The baryon density is given by $n=n_n+n_p$ and $\delta=(n_n-n_p)/n$ is the asymmetry,
with $n_n$ and $n_p$ denoting the neutron and proton densities, respectively.
This approach of Taylor expanding the energy functional up to fourth order around
the saturation density, $n_{0}$, has been applied recently in several
works \cite{Margueron2018a,Margueron2018b,Margueron2019,Ferreira:2019bgy}.
The empirical parameters can be identified as the coefficients of the expansion.  
The isoscalar empirical parameters are defined as successive density derivatives of $e_{\text{sat}}$,
$ P_{\text{is}}^{k}=(3n_{0})^k\partial^k e_{\text{sat}}/\partial n^k$ for symmetrical nuclear matter at saturation, i.e., at ($\delta=0,n=n_{0}$), while the isovector parameters are connected to the density derivatives of $e_{\text{sym}}$, $ P_{\text{iv}}^{k}=(3n_{0})^k\partial^k e_{\text{sym}}/\partial n^k$ also at ($\delta=0,n=n_{0}$).\\

Each EOS, denoted by index $i$, is represented by a point in the 8-dimensional space of parameters
\begin{align}
\text{EOS}_i &= (E_{\text{sym}},L_{\text{sym}},K_{\text{sat}},K_{\text{sym}},Q_{\text{sat}},Q_{\text{sym}},Z_{\text{sat}},Z_{\text{sym}})_i.
\end{align}
A set of EOS is generated by drawing random samples through  a
 multivariate Gaussian distribution
\begin{align}
\text{EOS}_i &\sim N(\boldsymbol{\mu},\boldsymbol{\Sigma}),
 \label{eq:Gauss_dis}
\end{align}
where the mean vector and (diagonal) covariance matrix are, respectively, 
$$\boldsymbol{\mu}^T=(\overline{E}_{\text{sym}},\overline{L}_{\text{sym}},\overline{K}_{\text{sat}},\overline{K}_{\text{sym}},\overline{Q}_{\text{sat}},\overline{Q}_{\text{sym}},\overline{Z}_{\text{sat}},\overline{Z}_{\text{sym}})$$ 
and
\begin{align}
\boldsymbol{\Sigma}=\text{diag}(\Sigma_{E_{\text{sym}}},&\Sigma_{L_{\text{sym}}},\Sigma_{K_{\text{sat}}},\Sigma_{K_{\text{sym}}}, \Sigma_{Q_{\text{sat}}},\Sigma_{Q_{\text{sym}}},\Sigma_{Z_{\text{sat}}},\Sigma_{Z_{\text{sym}}}).
\end{align}

While the lower order coefficients are quite well constrained experimentally \cite{Youngblood1999,Margueron2012,Li2013,Lattimer2013,Stone2014,OertelRMP16},  the higher order $Q_{\text{sat}},\, Z_{\text{sat}}$ and $K_{\text{sym}}, \, Q_{\text{sym}},
  \, Z_{\text{sym}}$ are poorly known
 \cite{Farine1997,De2015,Mondal2016,Margueron2018b,Malik2018,Zhang2018,Li2019}. 
In the present work, we fix  $E_{\text{sat}}=-15.8$ MeV  and $n_{0}=0.155$ fm$^{-3}$ as these quantities are rather well constrained. Table \ref{tab:parametros} shows the parameters means and standard deviations, $\sigma_{a}=\sqrt{\Sigma_a}$, we use \cite{Margueron2018b}.\\

\begin{table}[!htb]
	\begin{center}
		\begin{tabular}{lllllllll}
			\hline
			$P_{i}$  & $E_{\text{sym}}$ &  $L_{\text{sym}}$ & $K_{\text{sat}}$ & $K_{\text{sym}}$ & $Q_{\text{sat}}$ & $Q_{\text{sym}}$ & $Z_{\text{sat}}$ & $Z_{\text{sym}}$ \\
			\hline
			\hline
			$\overline{P}_{i}$ & $32$  & $60$ & $230$ & $-100$ & $300$ & $0$& $-500$ & $-500$ \\
			$\sigma_{P_i}$  & $2$ & $15$ & $20$ & $100$ & $400$ & $400$& $1000$ & $1000$\\
			\hline
		\end{tabular}
	\end{center}
	\caption{The means $\overline{P}_{i}$ and standard deviations
		$\sigma_{P_i}$ that characterize the multivariate Gaussian distribution [Eq. (\ref{eq:Gauss_dis})].
		 All the quantities are in units of MeV.}
	\label{tab:parametros}
\end{table}

Each valid EOS sampled describes $npe\mu$ matter in $\beta$-equilibrium and satisfy the following conditions
i) the pressure is an increasing function of density (thermodynamic stability); 
ii)  the speed of sound is smaller than the speed of light (causality);
iii) the EOS supports a maximum mass at least as high as $1.97M_{\odot}$
\cite{Arzoumanian2017,Fonseca2017,Demorest2010,Antoniadis2013}.
This EOS sampling approach introduces no a priori correlations between the parameters (zero covariances), and the correlations present in final set of valid EOS are solely induced from the above conditions. 
We use the nuclear SLy4 EOS \cite{Douchin2001} for the low density region of $n<0.5n_0$, and a power-law interpolation between $0.5n_0<n<n_0$ is performed to ensure a matching at $n_0$ with the sampled EOS  (different matching procedures were analyzed in \cite{Ferreira:2020wsf}).\\

Employing the EOS Taylor expansion around saturation for symmetric nuclear matter as the EOS parametrization itself is  widely used \cite{Margueron2018a,Margueron2018b,Zhang2018,Carson:2018xri,Tews2018,Xie2019,Margueron2019,Ferreira:2020kvu,Guven2020,Li2021a,Zhang2021}. There are no convergence issues when it is considered as an EOS parametrization, and this is a reliable approximation for realistic EOS around the saturation density for symmetric nuclear matter \cite{Margueron2018a}. In \cite{Margueron2018a} the authors have also shown that they could fit any EOS to a quite good accuracy using the EOS forth order Taylor expansion.  However, the higher-order parameters must be seen as effective terms, which may substantially deviate from the actual nuclear matter expansion coefficients. Nevertheless, the present EOS parametrization allows to directly constrain the EOS from the properties of symmetric nuclear matter and the symmetry energy near saturation density, while different EOS parametrizations impose such properties in a indirect way \cite{Zhang2018}. 
A recent work \cite{Biswas:2020puz} has pointed out that the use of such Taylor expansion parametrization may be an inadequate approach. The authors have compared the results from two parametrization procedures i) a third order Taylor expansion; ii) an hybrid formulation where a second order Taylor expansion describes
matter below $2n_0$ and three polytropes  are used to describe matter above
that density. While we agree that the present EOS parametrization has its limitations, it should not be compared with a hybrid procedure where polytropes are randomly added to the low density Taylor expansion, as it brakes the correlations between the information around the saturation density, from the Taylor expansion,  with the high-density region defined by the polytropes. In this approach, the correlations among the Taylor coefficients and the polytropes indices are  close to zero (which is visible in the posterior probabilities in \cite{Biswas:2020puz}), and both low- and high-density regions are disconnected. Since we are only interested in constraining the thermodynamic properties of the EOS and not the nuclear matter properties, and we consider a one branch EOS, the Taylor expansion is as  reliable as other parametrizations, describing smooth equations of state.  \\

For each EOS, we compute the dimensionless tidal deformability  
$\Lambda = \frac{2}{3}k_2C^{-5}$, where $k_2$ is the quadrupole tidal Love number and $C=GM/(c^2R)$ is the star's compactness \cite{Hinderer2008}. We also determine the effective tidal deformability of binary systems, 
\begin{equation}
\tilde{\Lambda}=\frac{16}{13}\frac{(12q+1)\Lambda_1+(12+q)q^4\Lambda_2}{(1+q)^5},
\label{eq:lambda}
\end{equation}
where  $q=M_2/M_1<1$ is the binary mass ratio and 
$\Lambda_{1}\, (M_1)$ and $\Lambda_{2}\, (M_2)$ represent the tidal deformability (mass) of the primary and the secondary NS in the binary, respectively. $\tilde{\Lambda}$ is the leading tidal parameter in gravitational-wave signal from a NS merger. The constraint $\tilde{\Lambda}=300^{+500}_{-190}$ (symmetric 90\% credible interval) was estimated from the GW170817 with a mass ratio  bounded as $0.73\leq q \leq1$ \cite{Abbott:2018wiz}. 
The binary chirp mass, $M_{\text{chirp}}=M_1 q^{3/5}/(1+q)^{1/5}$,
was measured as $M_{\text{chirp}}=1.186^{+0.001}_{-0.001}M_{\odot}$ for the GW170817 event \cite{Abbott:2018wiz}. In the present work, we fix the $M_{\text{chirp}}$ to $1.186M_{\odot}$. 

\section{Results}
\label{sec:results}
By applying the statistical procedure described in the last section, we have generated a dataset containing $150000$ EOS that describe neutron star matter in $\beta$-equilibrium.

\subsection{Subsets defined by $M_{\text{max}}$}
\begin{figure}[!htb]
	\centering
	\includegraphics[width=1.0\columnwidth]{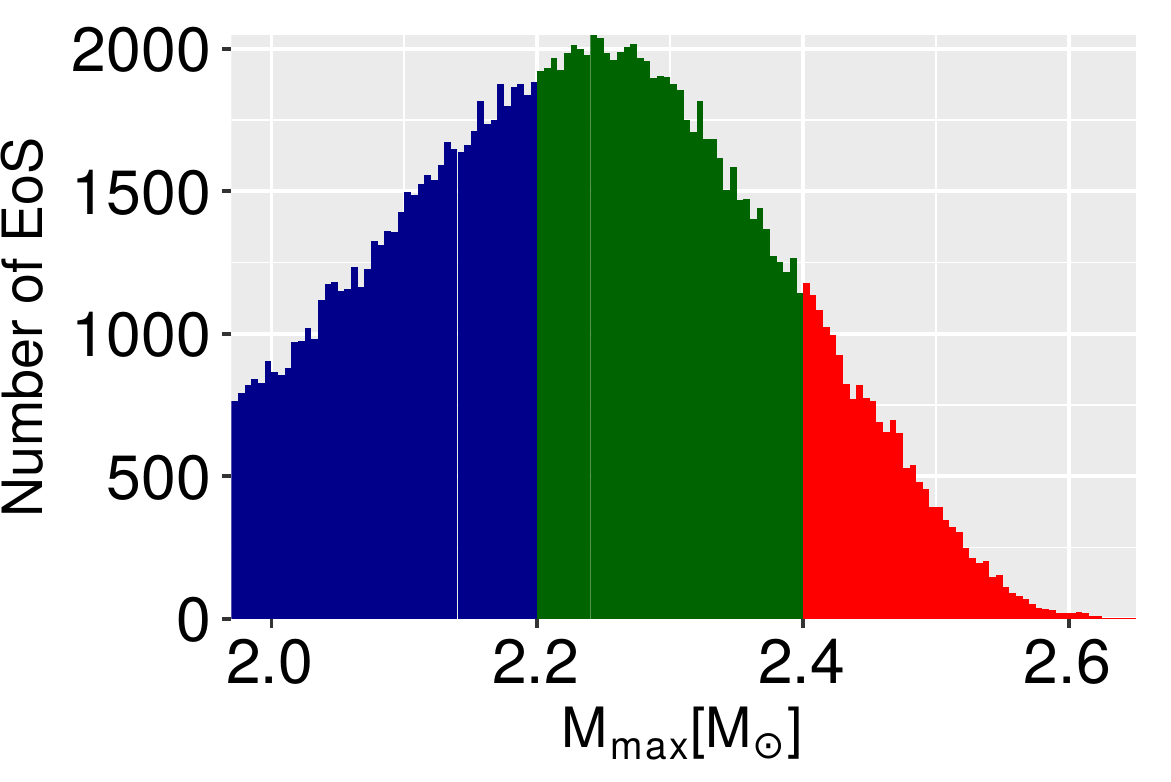}
	\caption{Number of EOS as a function of $M_{\text{max}}$. The colors identify the subsets: $1.97<M_{\text{max}}/M_{\odot} \leq2.20$ (blue),
	$2.20<M_{\text{max}}/M_{\odot} \leq2.40$ (green), 
	and $2.40<M_{\text{max}}/M_{\odot} \leq2.66$ (red). }
\label{fig:hist}
\end{figure}

We are interested in studying the effect of the maximum NS mass, $M_\text{max}$, on the NS EOS properties. Therefore,  we split our data set into three subsets depending on $M_{\text{max}}$ reached by each EOS: $1.97<M_{\text{max}}/M_{\odot}\leq 2.20$ (set 1), $2.20<M_{\text{max}}/M_{\odot}\leq 2.40$ (set 2), and $2.40<M_{\text{max}}/M_{\odot}\leq 2.66$ (set 3).
The highest value reached in our data-set is $M_{\text{max}}=2.66M_{\odot}$.
The histogram of the number of EOS as a function of $M_{\text{max}}$ is shown in Fig. \ref{fig:hist},
displaying an unimodal distribution with a peak around $2.25M_{\odot}$, being the most likely outcome for $M_{\text{max}}$ within the present EOS parametrization approach and sampling procedure.
Only $3162$ EOS were able to sustain a NS with $M_{\text{max}}>2.5 M_{\odot}$,
2.11\% of our dataset.\\

Let us recall that different experiments have recently concentrated on getting more information on the nuclear symmetry energy at saturation. In  \cite{Estee2021}, the authors have constrained  the symmetry energy  and its slope at saturation to, respectively, $32.5< E_{\text{sym}}< 38.1$ MeV and  $42<L_{\text{sym}}<110$ MeV  using  spectra of charged pions produced in intermediate energy collisions.  These values are in accordance with the ones obtained in \cite{Danielewicz2016} from charge exchange and elastic scattering reactions, respectively, $33.5< E_{\text{sym}}< 36.4$ MeV and   $70<L_{\text{sym}}<101$ MeV, but smaller than the ones determined from the measurement of the neutron radius of $^{208}$Pb by the PREX-2 collaboration \cite{PREX2,Reed2021}, which have obtained for the slope of the symmetry energy  $L_{\text{sym}}=106\pm37$ MeV.
These estimates are somehow larger than the ones proposed in \cite{Lattimer2013}, which take into account experimental, theoretical and observational constraints, e. g.  $29.0<E_{\text{sym}}<32.7$ MeV and $44<L_{\text{sym}}<66$ MeV (\cite{Lattimer2013} and updated in \cite{Lattimer2014}). The symmetry energy properties of our EOS, see  Table \ref{tab:param}, essentially reproduce at one standard deviation the constraints of \cite{Lattimer2014}, which are strongly influenced by chiral effective field theory results for neutron matter \cite{Hebeler2013}. However, the three sets also include EOS that fall within the recent experimental intervals  within two to three standard deviations. Moreover, the extremes are well out of these ranges.

Table \ref{tab:param} shows the mean and standard deviations of each set.  We see that, except for $Z_{\text{sym}}$, isovector properties decrease from set 1 to set 3. Larger $M_{\text{max}}$ values require smaller isovector properties, while the isoscalar properties show the opposite behaviour, e.g.,  $K_{\text{sat}}$ increases slightly and there is a considerable increase of $Q_{\text{sat}}$, but $Z_{\text{sat}}$ shows a  different behavior. Two comments are in order: i) both $E_{\text{sym}}$ and $L_{\text{sym}}$ are very similar in the three sets indicating that the maximum mass star is not sensitive to the symmetry energy close to saturation density; ii) since $Z_{\text{sat}}$ is the last term of the expansion considered, it has an effective character taking into account all the missed higher order terms.
As we mentioned in Sec. \ref{sec:EOS}, the high-order terms should be seen as effective ones, and their values might deviate from the true nuclear matter properties. A recent review \cite{Li2021a} discussing the constraints set on the nuclear symmetry energy by NS observations since the detection of GW170817, proposes for the  $L_{\text{sym}}$ and $K_{\text{sym}}$ values and uncertainties of the order of the ones presented in Table \ref{tab:param}.

\begin{table}[!htb]
\centering
\begin{tabular}{ccccccccc}
  \hline
&\multicolumn{2}{c}{\bf {Set 1}}& &\multicolumn{2}{c}{\bf {Set 2}} & &\multicolumn{2}{c}{ \bf {Set 3}}\\
\hline
$M_{\text{max}}/M_{\odot}$ & \multicolumn{2}{c}{$]1.97,2.20]$} & & \multicolumn{2}{c}{$]2.20,2.40]$}& &\multicolumn{2}{c}{$]2.40,2.66]$}\\
\# EOSs &\multicolumn{2}{c}{61660}& &\multicolumn{2}{c}{69789} & & \multicolumn{2}{c}{18551}\\
\cline{2-3} \cline{5-6} \cline{8-9}
 & mean & std & & mean & std & & mean & std \\ 
  \hline
  $K_{\text{sat}}$ & 229.28 & 19.83   & &  233.17 & 19.63    & & 238.74 & 19.75    \\ 
  $Q_{\text{sat}}$ & -93.44 & 65.23   & &  18.11 & 71.09     & & 148.11 & 77.45   \\ 
  $Z_{\text{sat}}$ & -3.47 & 67.31    & &  -109.11 & 81.75   & & -237.37 & 96.68   \\ \ 
  $E_{\text{sym}}$ & 32.01 & 1.98     & &  32.01 & 2.00      & & 31.90 & 2.01     \\ 
  $L_{\text{sym}}$ & 61.21 & 14.85    & &  60.97 & 14.89     & & 58.19 & 14.89     \\ 
  $K_{\text{sym}}$ & -64.06 & 90.62   & &  -72.54 & 89.88    & & -107.78 & 78.87   \\ 
  $Q_{\text{sym}}$ & 300.44 & 328.93  & &  227.27 & 320.31   & & 112.70 & 277.83   \\  
  $Z_{\text{sym}}$ & 392.77 & 684.98  & &  292.95 & 598.48   & & 433.28 & 509.64   \\  
   \hline
\end{tabular}
\caption{The mean [MeV] and standard deviation [MeV] (std) of the EOS parameters for each set. In the following figures the different sets are represented by the colors blue (Set 1), green (Set 2) and red (Set 3). }
\label{tab:param}
\end{table}

\subsubsection{Neutron stars properties}

\begin{figure*}[!htb]
	\centering
	\includegraphics[width=0.9\columnwidth]{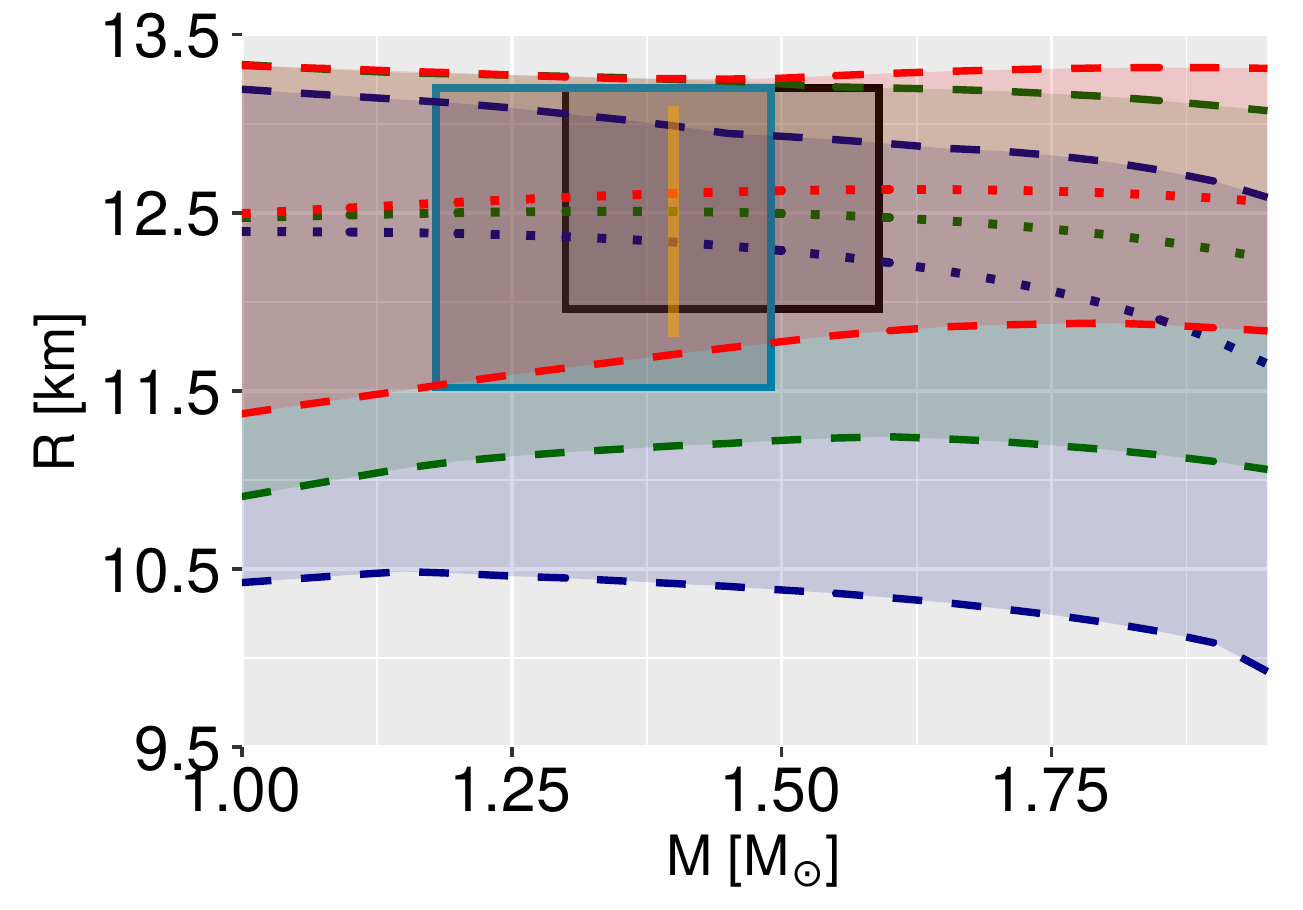}
	\includegraphics[width=0.9\columnwidth]{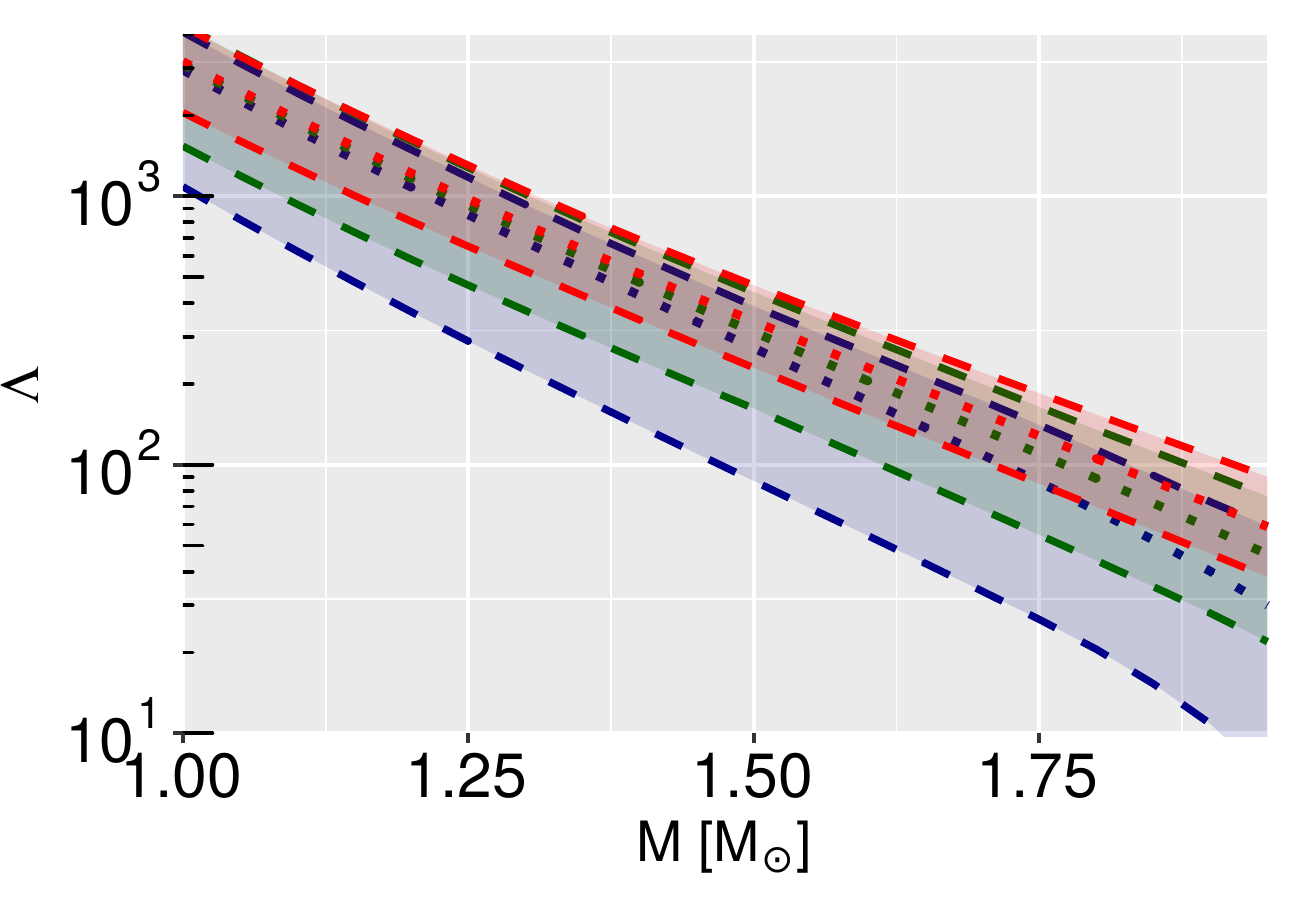}
	\caption{The minimum/maximum (dashed lines) and mean (dotted lines) values for the radius (left panel) and tidal deformability (right panel) as a function of NS mass. The color indicates the set: set 1 (blue), set 2 (green), and set 3 (red). The regions enclosed by solid lines display $(M,R)$ constraints obtained by two independent analysis using the NICER x-ray data from the millisecond pulsar PSR J0030+0451 \cite{Riley_2019,Miller:2019cac}, while the orange bar shows the recent constraint, $R_{1.4M_\odot}=12.45\pm 0.65$ km, extracted from a combination of NICER and XMM-Newton data \cite{Miller2021}.    }
\label{fig:R_Lambda_as_f_M}
\end{figure*}

In order to study NS properties, we integrate the Tolmann-Oppenheimer-Volkoff (TOV)
equations \cite{TOV1,TOV2}, together with the differential equations that determine
the tidal deformability \cite{Hinderer:2009ca}. 
Figure \ref{fig:R_Lambda_as_f_M} shows the radius (left panel) and tidal deformability (right panel) as a function of the NS mass for each set. Statistics for some specific NS masses are given in Table \ref{tab:stats}. As expected, larger $M_{\text{max}}$ values correspond to larger NS radii and tidal deformabilities. The lower bound on $R$ is more sensitive to $M_{\text{max}}$ than the upper bound.  
However, we are unable to constrain the $M_{\text{max}}$ value from the analysis of the millisecond pulsar PSR J0030+0451 (obtained from the NICER x-ray data) \cite{Riley_2019,Miller:2019cac} since all sets describe the NICER data. Next, we will also conclude that we cannot constrain the $M_{\text{max}}$ from the tidal deformability inferred from the GW170817 event \cite{Abbott:2018wiz}.

\begin{figure*}[!htb]
	\centering
	\includegraphics[width=0.9\columnwidth]{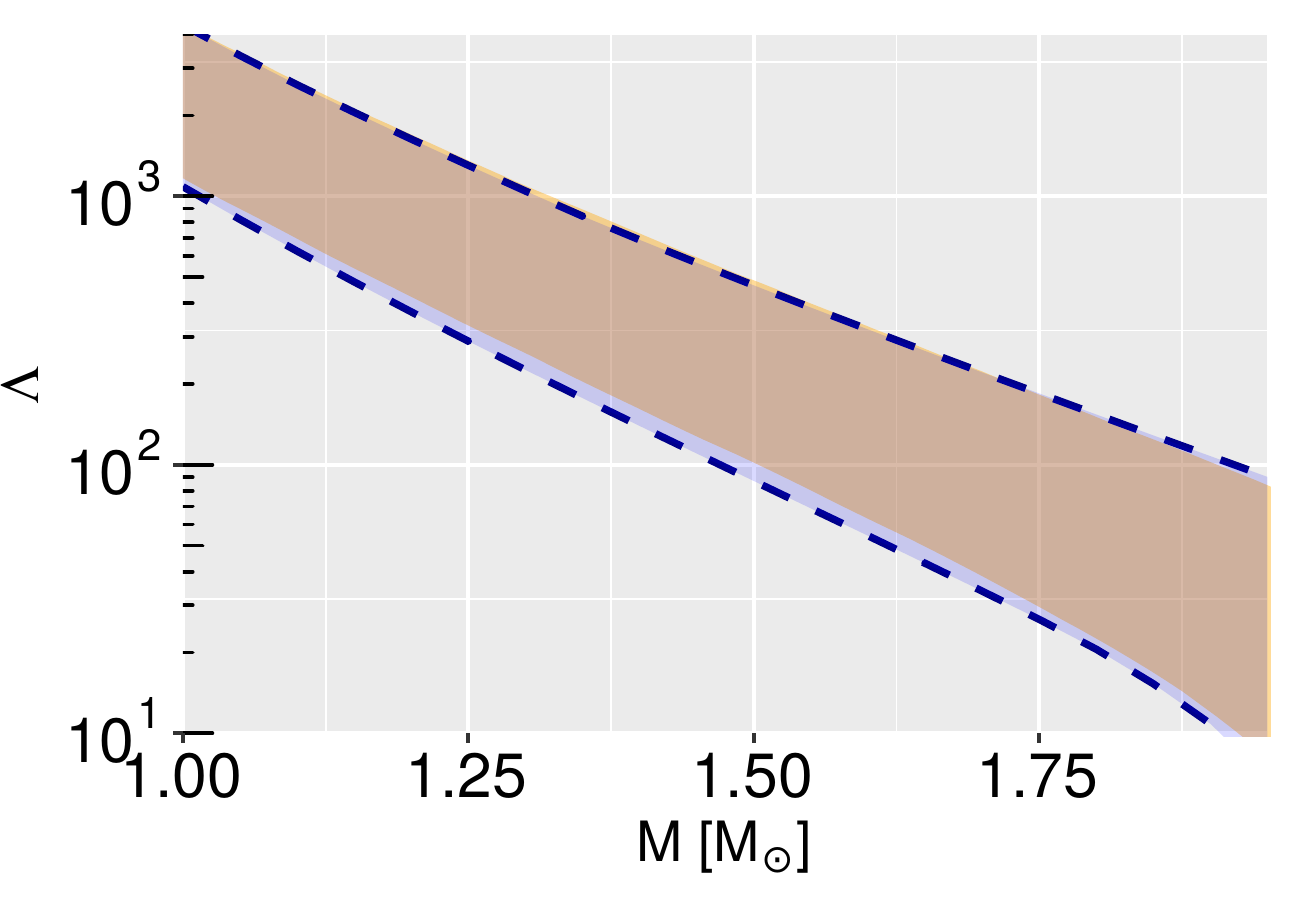}
	\includegraphics[width=0.9\columnwidth]{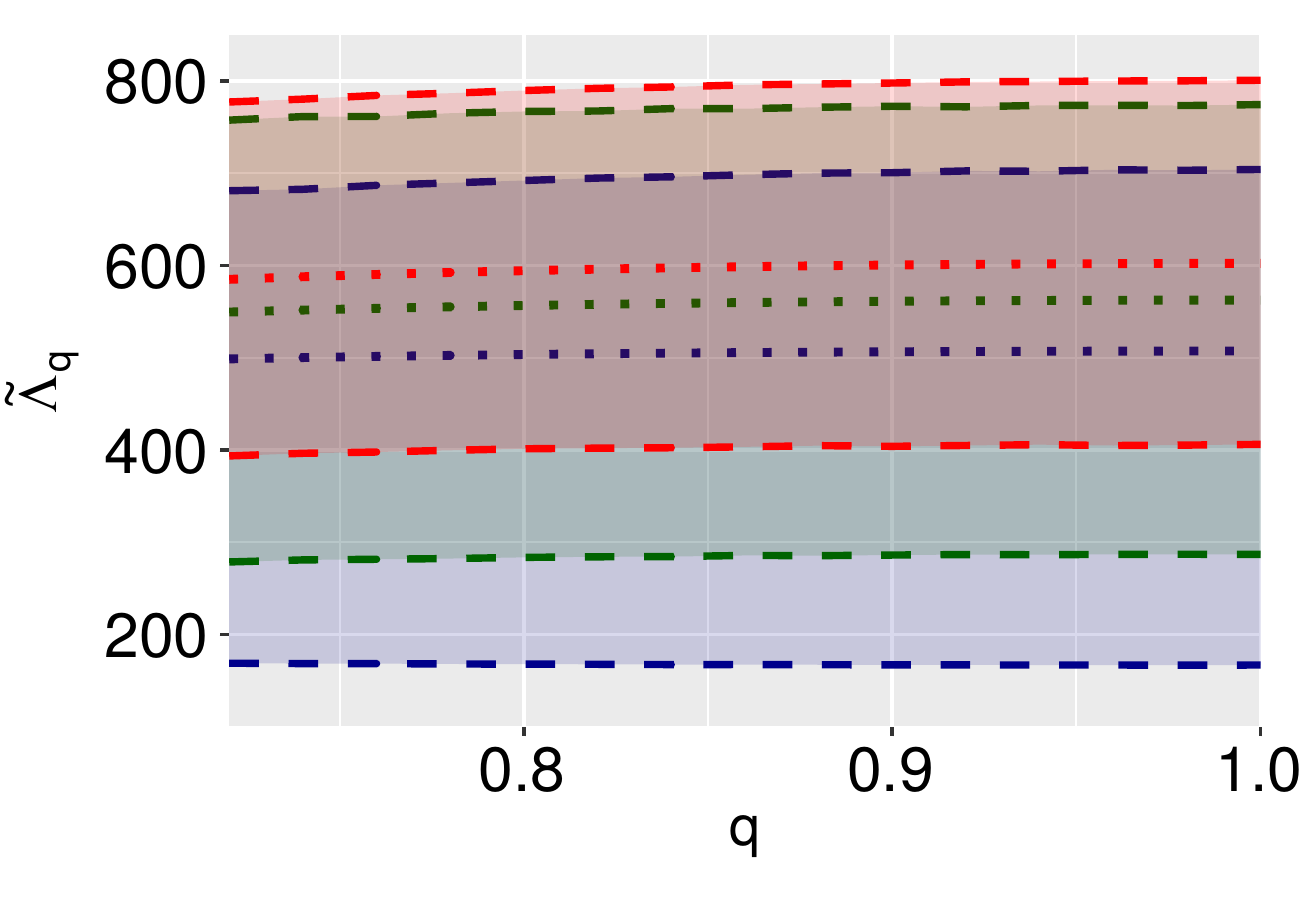}
	\caption{Left panel: the tidal deformability minimum/maximum values (dashed lines) for the entire data set, i.e., $1.97 <M_{\text{max}}/M_{\odot} \leq2.66$. The red region corresponds to
	 the 90\% posterior credible level obtained in \cite{Abbott:2018exr}, when it is imposed that the EOS should describe 1.97$M_\odot$ NS.; Right panel: the minimum/maximum (dashed lines) and mean (dotted lines) values for the effective tidal deformability of the binary, $\tilde{\Lambda}$, as a function of $q$ with $M_{\text{chirp}}=1.186M_{\odot}$. The color indicates the set: set 1 (blue), set 2 (green), and set 3 (red). }
	\label{fig:binaries_tidal}
\end{figure*}

Figure \ref{fig:binaries_tidal} shows the extreme values (maximum and minimum) for the tidal deformability from our entire data set (left panel) and the effective tidal deformability of the binary ($M_{\text{chirp}}=1.186M_{\odot}$) as a function of the binary mass ratio $q=M_2/M_1$. Two main conclusion can be extracted: i) the range of $\tilde{\Lambda}$ shifts to higher values as $M_{\text{max}}$ increases, but all sets are compatible with 
$\tilde{\Lambda}=300^{+500}_{-190}$ (symmetric 90\% credible interval) from the GW170817 event \cite{Abbott:2018wiz}; ii) furthermore, the NS tidal deformability range of our data set is consistent with the 90\% posterior credible level obtained in \cite{Abbott:2018exr}, when the analysis assumes that the EOS should describe 1.97$M_\odot$ NS. 
Some statistics for the different sets are given in Table \ref{tab:stats}.

\begin{table*}[!htb]
\centering
\begin{tabular}{ccccccccccccccccccc}
  \hline
  & \multicolumn{4}{c}{\bf Set 1 } 
  & & \multicolumn{4}{c}{\bf Set 2}
  &  &\multicolumn{4}{c}{\bf Set 3}\\
    & \multicolumn{4}{c}{$1.97<M_{\text{max}}/M_{\odot}\leq 2.20$ } 
  & & \multicolumn{4}{c}{$2.20<M_{\text{max}}/M_{\odot}\leq 2.40$ }
  &  &\multicolumn{4}{c}{$2.40<M_{\text{max}}/M_{\odot}\leq 2.66$ }\\
  \cline{2-5}  \cline{7-10}  \cline{12-15}
&  mean & std  & min&  max & & mean & std  & min&  max & & mean & std  & min&  max   \\
  \hline
$R_{1.4M_{\odot}}$ [km]  &  $12.34$ & 0.22  & 10.42 & 12.99 & & 12.51 & 0.22  & 11.19 &  13.25 & & 12.61 &  0.20  & 11.70&  13.25  \\
$R_{1.6M_{\odot}}$ [km]  &  $12.22$ & 0.24  & 10.34 & 12.89 & & 12.47 & 0.21  & 11.24 &  13.20 & & 12.63 &  0.20  & 11.84&  13.28  \\
$R_{2.08M_{\odot}}$ [km]  &  $11.37$ & 0.36  &  9.76 &  12.36 & & 12.05 & 0.26  &  10.89&   12.95 & & 12.48 &  0.19  & 11.76 & 13.29 \\
$\Lambda_{1.4M_{\odot}}$   &  $427.36$ & 50.37  & 139.55 & 598.21 & & 477.83 & 50.23  & 245.1 &  660.47 & & 514.68 & 47.29 & 346.4&  688.17  \\
$\Lambda_{1.6M_{\odot}}$   &  $260.01$ & 25.05  & 54.45 & 260.01 & & 205.37 & 23.56  & 105.18 &  296.23 & & 229.95 & 21.95 & 153.81 &  319.37  \\
$\tilde{\Lambda}_{q=0.9}$  &  $506.41$ & 57.53  & 167.28 & 700.38 & & 561.13 & 58.07  & 286.23 &  772.35 & & 600.42 & 54.83   & 403.99&  797.54 \\
$n_{\text{max}}/n_0$      &  7.09 & 0.42  & 5.72 & 8.65 & & 6.35 & 0.27  & 5.51 &  7.32 & & 5.76 & 0.16  & 5.05 &  6.19\\
$(v_s^{\text{cen}})^2$ [$c^2$]      &  $0.75$ & 0.17  & 0.03 & 1 & & 0.86 & 0.12  & 0.02 &  1 & & 0.91 & 0.08  & 0.36&  1  \\
   \hline
\end{tabular}
	\caption{Sample statistics for $R_{M_i}$, $\Lambda_{M_i}$, $\tilde{\Lambda}_{q=0.9}$, and the NS' central density $n_{\text{max}}/n_0$ and speed of sound squared $(v_s^{\text{cen}})^2$ for the three sets considered.}
	\label{tab:stats}
\end{table*}

Notice that in \cite{Raaijmakers2021} the authors have obtained for the radius of a 1.4 $M_\odot$ star 
12.33$^{+0.76}_{-0.81}$km  and 12.18$^{+0.56}_{-0.79}$ km  within two different models. The radii obtained within our sets, see Table \ref{tab:stats}, are in accordance with these predictions. However, the prediction obtained for the radius of  the PSR J0740+6620 in \cite{Miller2021}, $12.39^{+1.30}_{-0.98}$ km with the mass $2.072 ^{+0.067}_{-0.066} M_\odot$ \cite{Riley2021} seems to be more compatible with the two sets with larger maximum masses, although set 1 is not excluded. 

We also give the prediction for the radius of a 1.6$M_\odot$ star that has been constrained in \cite{Bauswein2017} to be larger than 10.68$^{+0.15}_{-0.04}$km considering that there was no prompt collapse following GW170817 as suggested by the associated electromagnetic emission. All our three sets satisfy this constraint.  As expected, the predicted  central density of maximum mass stars decreases from the less massive set to the most massive. For the last set we have obtained a central density of $5.05<n_{\text{max}}/n_0< 6.19$ compatible with the values proposed in \cite{Legred2021} for cold non-spinning NSs.

\subsubsection{EOS thermodynamics}

\begin{figure*}[!htb]
	\centering
	\includegraphics[width=0.9\columnwidth]{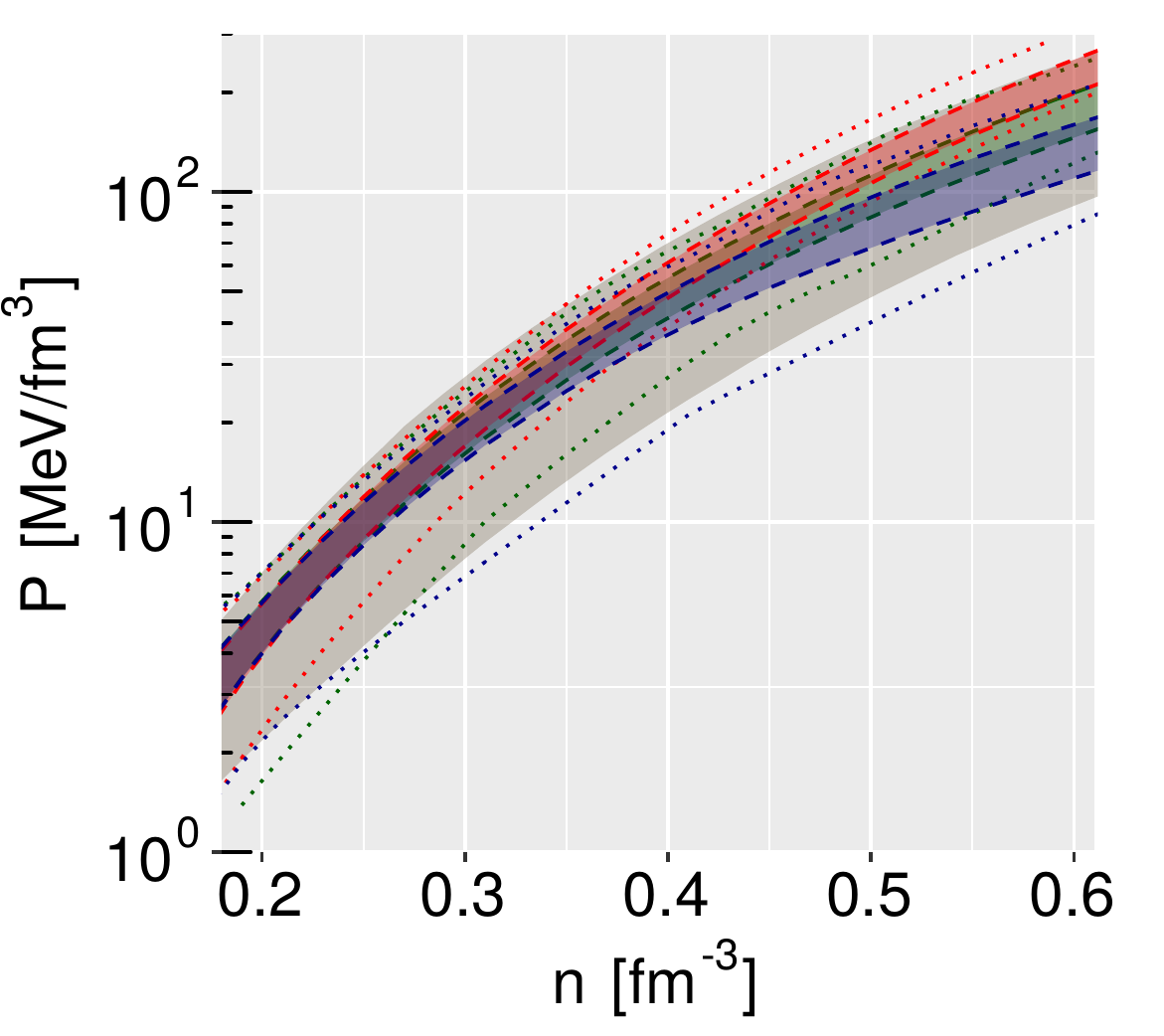}
    \includegraphics[width=0.9\columnwidth]{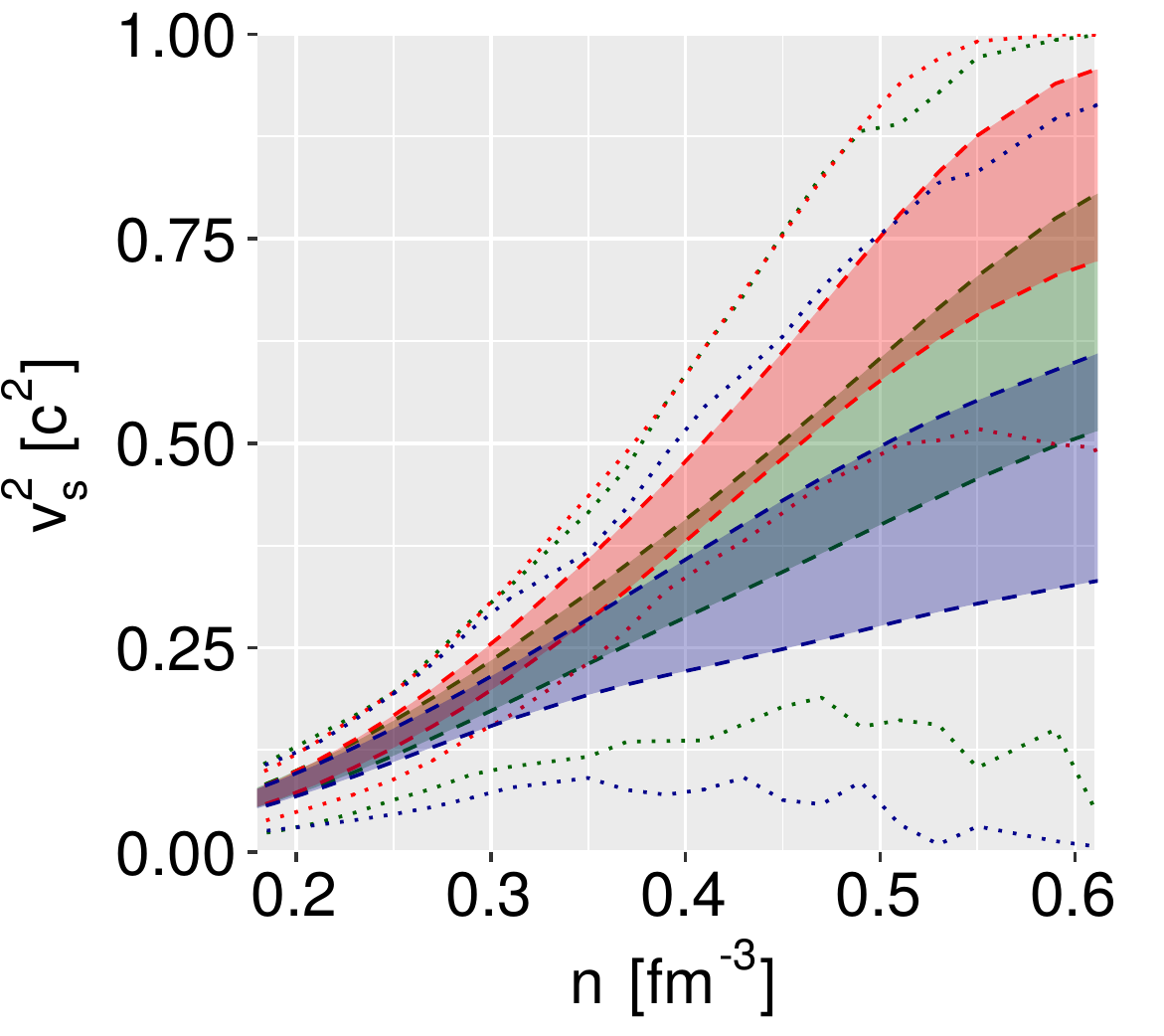}
	\caption{The 90\% credible levels (dashed lines) and maximum/minimum values (dotted lines) for the pressure (left) and speed of sound squared (right) as a function of baryonic density. 
	The color identifies the set: set 1 (blue), set 2 (green), and set 3 (red). 
	The gray region in $p(n)$ corresponds
to the 90\%  posterior credible level from \cite{Abbott18}.}
\label{fig:sample_pressure_vs}
\end{figure*}
Let us now analyze the thermodynamics properties of each EOS subset. 
The pressure and speed of sound are shown in Fig. \ref{fig:sample_pressure_vs}, where the
shaded colored regions represent the 90\% credible (equal-tailed) interval regions (sets 1, 2 and 3 are respectively, blue, green and red). For the pressure (left panel), the results of each set begins to deviate from each other only at moderate densities, $n\approx0.3$ fm$^{-3}$. Although the results are within the range of the LIGO/Virgo analysis (gray region), there is a considerable discrepancy for densities $n<0.55$ fm$^{-3}$, although the extreme values cover a larger region. 
This is expected since the EOS parametrization we use is constrained at low densities by the low order nuclear matter parameters, such as $L_{\text{sym}}$ and $K_{\text{sat}}$, while a polytropic approach as the one used in the analysis of GW170917, being solely constrained by thermodynamic stability conditions, is able to reproduce extreme behaviours at any density region (even close to the saturation density).  

The speed of sound (right panel of Fig. \ref{fig:sample_pressure_vs}) also shows a similar dependence at low densities for all sets, but a different prediction at higher densities. If we compare set 1 (blue), $1.97<M_{\text{max}}/M_{\odot}<2.20$, with set 3 (red), $2.40<M_{\text{max}}/M_{\odot}<2.66$, we see that the regions do not overlap for $n>0.4$ fm$^{-3}$, showing that the high density dependence of $v_s^2$ is considerable bound to the $M_{\text{max}}$ value. Moreover, the extremes show the possibility of occurring a large softening of the EOS, as would be present for a phase transition. These are the EOS that in the present approach cover the lower range of the $P(n)$ band determined by LIGO/Virgo collaboration. On the other hand, the upper limit of the speed of sound is clearly defined by the causality constraint at high densities.

\subsection{Constraining the EOS from $\tilde{\Lambda}$}

The analysis of the GW170817 event by the LIGO/Virgo collaboration has provided a constraint on the effective tidal deformability of the binary system, $\tilde{\Lambda}=300^{+500}_{-190}$ (symmetric 90\% credible interval) \cite{Abbott:2018wiz}. A possible way of learning how such a credible region constrains the properties of the nuclear matter consists in filtering out all EOS that do not fulfill such interval and analyze the properties of the remaining EOS. However, the set of EOS  used in the present study predicts $\tilde{\Lambda}$ values that range between $168$ and $790$ (sample minimum/maximum values), not reproducing neither $110<\tilde{\Lambda}<168$ nor $\tilde{\Lambda}>790$, i.e.,  the smallest and largest values obtained from the GW170817 analysis are not represented in our set. Notice, however, that the  kilonova/macronova AT2017gfo sets a  lower
bound on binary tidal deformability, which should satisfy, for the  GW170817 event, $\tilde \Lambda_{\rm}>200$ according to \cite{Kiuchi2019} or  $\tilde \Lambda_{\rm}
>300$  according to \cite{Radice2017,Radice2018}, and, therefore, our lower bound is reasonable.  Even though our set statistics presents a good agreement with  $\tilde{\Lambda}=300^{+500}_{-190}$ (see Table \ref{tab:stats}), in the following we explore a probabilistic inference of the EOS using the entire information encoded in the probability distribution of $\tilde{\Lambda}$ for the GW170817 event \cite{Abbott:2018wiz} and a multivariate Gaussian distribution of the EOS based on the properties of our EOS sets. Note that the formalism developed in the present section considers the entire dataset, without any class subset related with the maximum mass $M_{\text{max}}$.\\

\subsubsection{Probability distributions for the pressure and speed of sound}
We use a joint multivariate Gaussian distribution and its conditioning properties to
constrain the EOS thermodynamics, namely $p(n)$ and $v_s^2(n)$. This allows us to consider the complete information encoded in the posterior probability density function (pdf) of $\tilde{\Lambda}$,  $P_{LV}(\tilde{\Lambda})$ \cite{Abbott:2018wiz}.\\

Let us consider a $N$-multivariate Gaussian probability distribution $P(z_1,z_2, ... , z_N)={\cal N}(z_1,z_2,...z_N;\boldsymbol \mu, \boldsymbol\Sigma)$, where $\boldsymbol \mu=(E[z_1],E[z_2],...E[z_N])^T$ represents the mean vector and $\Sigma_{ij}=\operatorname {E} [(z_{i}-\mu _{i})(z_{j}-\mu _{j})]$ are the covariance matrix entries ($E$ represents the expected value operator).  To simplify the notation, let us rewrite the above probability distribution as 
\begin{equation}
    {\cal N}(z_1,z_2,...z_N;\boldsymbol \mu, \boldsymbol\Sigma)={\cal N}(\mathbf{X},\mathbf{Y};\boldsymbol \mu, \boldsymbol\Sigma),
    \label{eq:gaussian}
\end{equation}
where $\mathbf{X}=(z_1, ... , z_q)^T$ and $\mathbf{Y}=(z_{q+1}, ... , z_N)^T$ are, respectively, two subsets with dimension $q$ and $N-q$ of the original $N$ random variables. This way, the mean vector and covariance matrix are partitioned into blocks,
\begin{equation}
 \boldsymbol \mu=\begin{pmatrix}
  E[\mathbf{X} ] \\
  E[\mathbf{Y} ] \\
\end{pmatrix}\equiv
\begin{pmatrix}
  \boldsymbol \mu_{\mathbf{X}} \\
  \boldsymbol \mu_{\mathbf{Y}} \\
\end{pmatrix}
\quad \quad \text{and} \quad \quad 
\boldsymbol\Sigma=
\begin{pmatrix}
  \boldsymbol \Sigma_{\mathbf{X}\mathbf{X}} & \boldsymbol\Sigma_{\mathbf{X}\mathbf{Y}}\\
  \boldsymbol \Sigma_{\mathbf{Y}\mathbf{X}} & \boldsymbol \Sigma_{\mathbf{Y}\mathbf{Y}}\\
\end{pmatrix},
\label{eq:gaussian_dist}
\end{equation}
respectively. For instance, the block $\boldsymbol \Sigma_{\mathbf{X}\mathbf{Y}}$ is a $q\times (N-q)$ matrix with the $(1,1)$ entry being $\operatorname {E} [(z_{1}-\mu _{1})(z_{q+1}-\mu _{q+1})]$.\\

A crucial operation for Gaussian distributions is the conditional probability. It allows us to  determine the probability of a set of variables depending (conditioning) on the other set variables,  yielding a modified - but still - Gaussian probability distribution. 
Conditioning the pdf $P(\mathbf{X},\mathbf{Y})={\cal N}(\mathbf{X},\mathbf{Y};\boldsymbol \mu, \boldsymbol\Sigma)$ upon 
the set variables $\mathbf{Y}$ gives 
\begin{equation}
P(\mathbf{X}|\mathbf{Y})={\cal N}(\mathbf{X}| \mathbf{Y};\bar{\boldsymbol \mu},  \bar{\boldsymbol\Sigma}),
\label{eq:0}
\end{equation}
where 
\begin{align}
\bar{\boldsymbol\mu}
&=
\boldsymbol\mu_{\mathbf{X}} + \boldsymbol\Sigma_{\mathbf{X}\mathbf{Y}} \boldsymbol\Sigma_{\mathbf{Y}\mathbf{Y}}^{-1}
\left( \mathbf{Y} - \boldsymbol\mu_{\mathbf{Y}} \right)\label{eq:1}\\
{\overline {\boldsymbol\Sigma }}&={\boldsymbol\Sigma }_{\mathbf{X}\mathbf{X}}-{\boldsymbol\Sigma}_{\mathbf{X}\mathbf{Y}}{\boldsymbol\Sigma }_{\mathbf{Y}\mathbf{Y}}^{-1}{\boldsymbol\Sigma}_{\mathbf{Y}\mathbf{X}}.
\label{eq:2}
\end{align}
The dimensions of the mean vector, $\bar{\boldsymbol\mu}$, and  of the covariance matrix, ${\overline {\boldsymbol {\Sigma }}}$, are $q\times1$ and $q\times q$, respectively. \\

Our goal is to infer the thermodynamic properties of the EOS from the knowledge of $\tilde{\Lambda}$, extracted from the GW170817 event. Given the current uncertainty on the $M_{\text{max}}$, we want to analyze how the EOS inference depends on the $M_{\text{max}}$ value. In the following, we consider the EOS pressure, $p(n)$, and speed of sound, $v_s^2(n)$. For each density $n$, we model the probability distributions $P(p(n), M_{\text{max}}, \tilde{\Lambda})$  and $P(v_s^2(n), M_{\text{max}}, \tilde{\Lambda})$ as multivariate Gaussian distributions (Eq. \ref{eq:gaussian}), where both 
the mean vector and covariance matrix have a density dependence.
Then, we condition upon $\tilde{\Lambda}$, i.e., $P(p(n), M_{\text{max}} | \tilde{\Lambda})$
and $P(v_s^2(n), M_{\text{max}} | \tilde{\Lambda})$ by applying Eq. (\ref{eq:0}),  
where $\mathbf{X}$ corresponds to $(p(n), M_{\text{max}})^T$ and  $(v_s^2(n), M_{\text{max}})^T$, respectively, whereas $\mathbf{Y}=(\tilde{\Lambda})$.  Next, we determine the joint probability distributions  $P(p(n), M_{\text{max}})$ and $P(v_s^2(n), M_{\text{max}})$, by marginalizing over $\tilde{\Lambda}$, weighting by the prior pdf $P_{LV}(\tilde{\Lambda})$ \cite{Abbott:2018wiz},

\begin{align}
    P(p(n),M_{\text{max}})&=\int P(p(n),M_{\text{max}}|\tilde{\Lambda})P_{LV}(\tilde{\Lambda}) d\tilde{\Lambda}.
    \label{eq:cond_1}\\
    P(v_s^2(n),M_{\text{max}})&=\int P(v_s^2(n),M_{\text{max}}|\tilde{\Lambda})P_{LV}(\tilde{\Lambda}) d\tilde{\Lambda}.
    \label{eq:cond_2}
\end{align}
The probability distribution $P(p(n),M_{\text{max}})$ is no longer Gaussian, due to the non-Gaussian characteristic of the prior $P_{LV}(\tilde{\Lambda})$. To obtain the probability distribution for the pressure, given a specific value of $M_{\text{max}}$, say $M_{\text{max}}=a$, we calculate $P(p(n)|M_{\text{max}}=a)=P(p(n),M_{\text{max}}=a)/P(M_{\text{max}}=a)$.
Finally, we characterize this pdf using symmetric 90\%  credible intervals.
The same procedure is applied for the speed of sound.
Equations (\ref{eq:cond_1})-(\ref{eq:cond_2}) can be interpreted as compound pdfs. 
Note that any pdf can be thought as a marginal of some (higher dimension) joint pdf, e.g.,  $P(p(n),M_{\text{max}})=\int P(p(n),M_{\text{max}},\tilde{\Lambda})d\tilde{\Lambda}$, and, furthermore, any joint pdf can be written as a conditional pdf, e.g.,  $P(p(n),M_{\text{max}},\tilde{\Lambda})=P(p(n),M_{\text{max}}|\tilde{\Lambda})P(\tilde{\Lambda})$. That is, Eqs. (\ref{eq:cond_1})-(\ref{eq:cond_2}) become trivial identities if we change $P_{LV}(\tilde{\Lambda})$ by the (dataset) marginal distribution $P(\tilde{\Lambda})$. 
Compound pdfs arise when one uses an "external" pdf for some of the probabilistic model variables, i.e., 
we are integrating out $\tilde{\Lambda}$ from our probabilistic models $P(p(n),M_{\text{max}}|\tilde{\Lambda})$ and $P(v_s^2(n),M_{\text{max}}|\tilde{\Lambda})$ by weighting each value of $\tilde{\Lambda}$ by a specific (prior) pdf, 
which has a different structure than the datatset $P(\tilde{\Lambda})$.\\

\begin{figure*}[!htb]
	\centering
	\includegraphics[width=0.9\columnwidth]{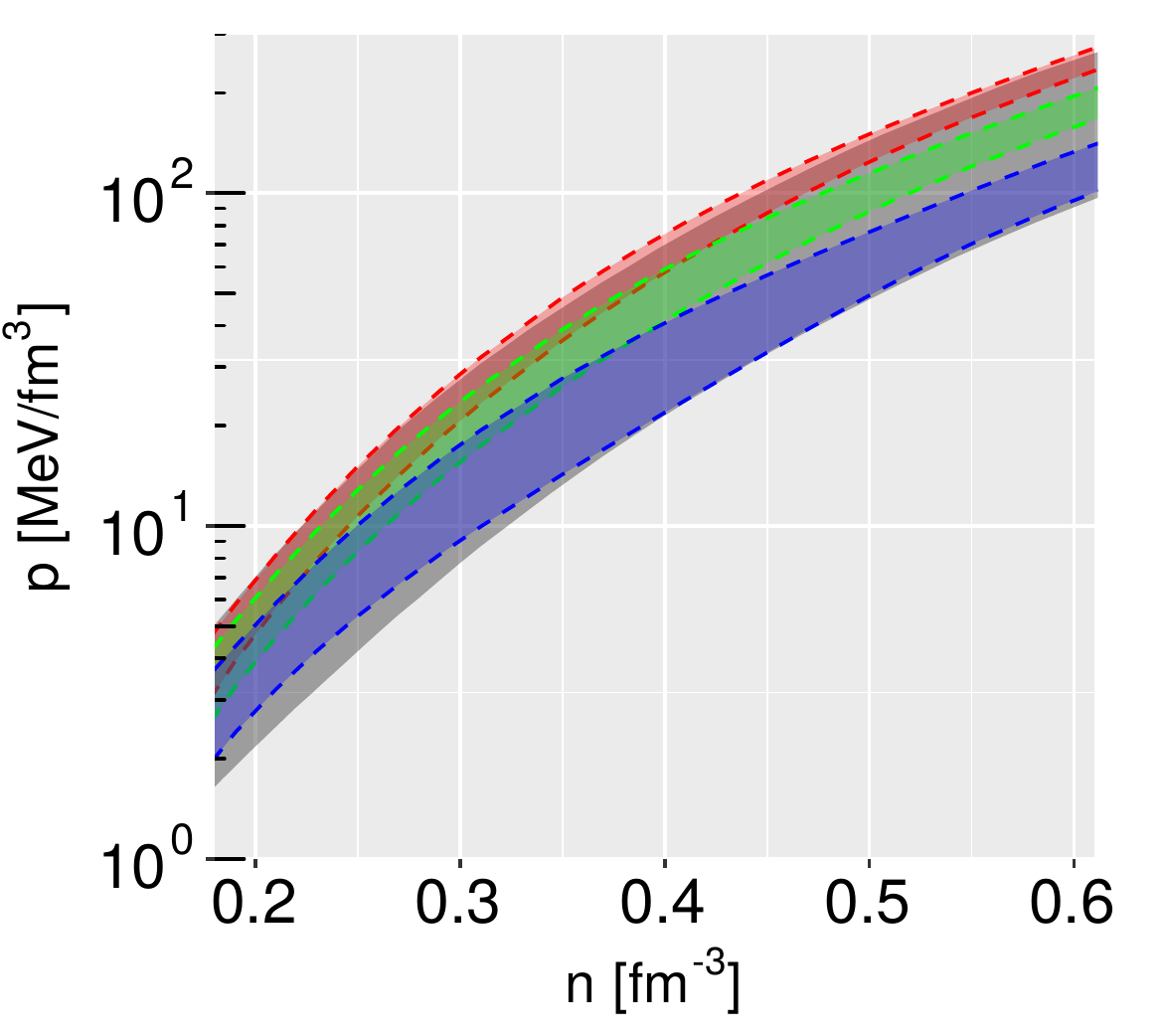}
	\includegraphics[width=0.9\columnwidth]{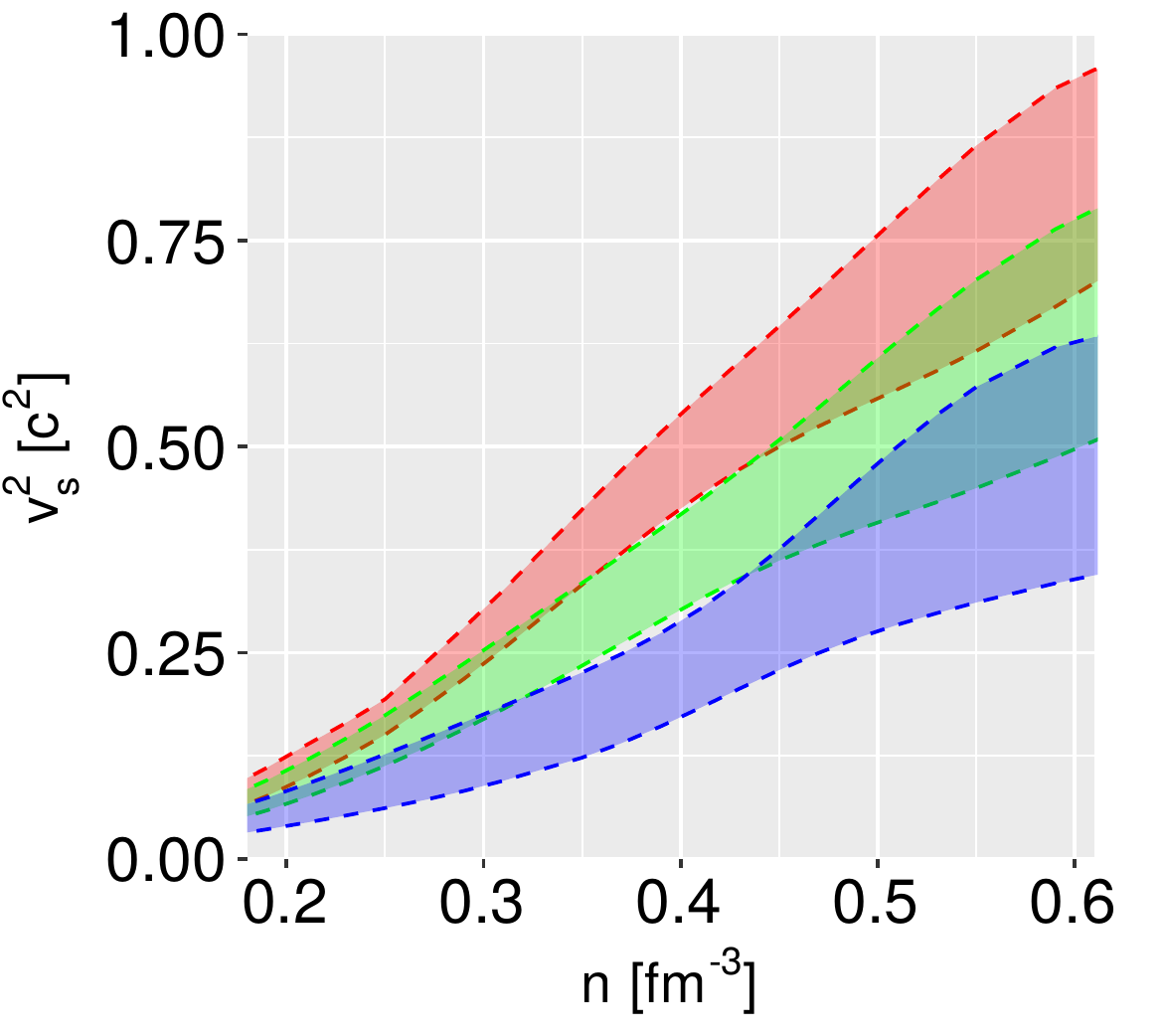}
	\caption{Inference results for the pressure (left) and speed of sound squared (right) as a function of baryonic density. The color regions show the 90\%  credible intervals for the pdfs $P(p(n)|M_{\text{max}})$ (left) and $P(v_s^2(n)|M_{\text{max}})$ (right) for three values of $M_{\text{max}}/M_{\odot}$: $2.0$ (blue), $2.3$ (green), and $2.6$ (red).The gray region in the $p(n)$ plot corresponds to the 90\%  posterior credible level from \cite{Abbott18}.}
\label{fig:inference1}
\end{figure*}

Figure \ref{fig:inference1} shows the results for $P(p(n)|M_{\text{max}})$ (left) and $P(v_s^2(n)|M_{\text{max}})$ (right), for three values of $M_{\text{max}}/M_{\odot}$: $2.0$ (red), $2.3$ (green), and $2.6$ (blue). The gray band on the $P(n)$ plot indicates the prediction region determined by the LIGO/Virgo analysis \cite{Abbott18}. It is interesting to conclude that: i) the bands do not overlap, i.e. knowing the maximum star mass it is possible to extract quite constrained information on the high density density EOS; ii) the three bands obtained with different maximum  masses lie almost inside the 90\%  credible interval predicted  from the GW170817 by LIGO/Virgo, obtained imposing that a maximum star mass equal to 1.97 $M_\odot$, the main difference lying on the upper boundary corresponding to the more massive stars, iii) the probabilistic approach which allows to take into account the whole GW170817 pdf has given more freedom to the low density EOS, giving rise to distinct bands also at low densities contrary to the previous conclusions. The $p(n)$ band obtained by the LIGO/Virgo collaboration was obtained from a set of EOS not constrained by nuclear matter properties and, therefore, may cover a region that would be excluded by nuclear properties. 

Using the probabilistic model, it was possible to determine the expected pressure band for a maximum star mass within a well defined mass range. For mass intervals that take as the lower limit observed 2 solar mass stars the lower limit of the LIGO/Virgo collaboration pressure band is not populated at 90\% credible probability. For the speed of sound, a monotonic increasing function of the density, values quite far from the conformal limit were obtained. The  maximum mass clearly constrains the speed of sound, with values close to one being obtained only if a maximum mass of 2.66 $M_\odot$ is imposed. The main effect of the probabilistic approach that takes into account the full pdf obtained by the LIGO/Virgo collaboration for the tidal deformability with respect to the study done in the previous section is to enlarge the speed of sound range at low densities, allowing for quite low values of the speed of sound, and, therefore giving rise to a harder EOS at intermediate densities. \\

\begin{figure*}[!htb]
	\centering
	\includegraphics[width=0.9\columnwidth]{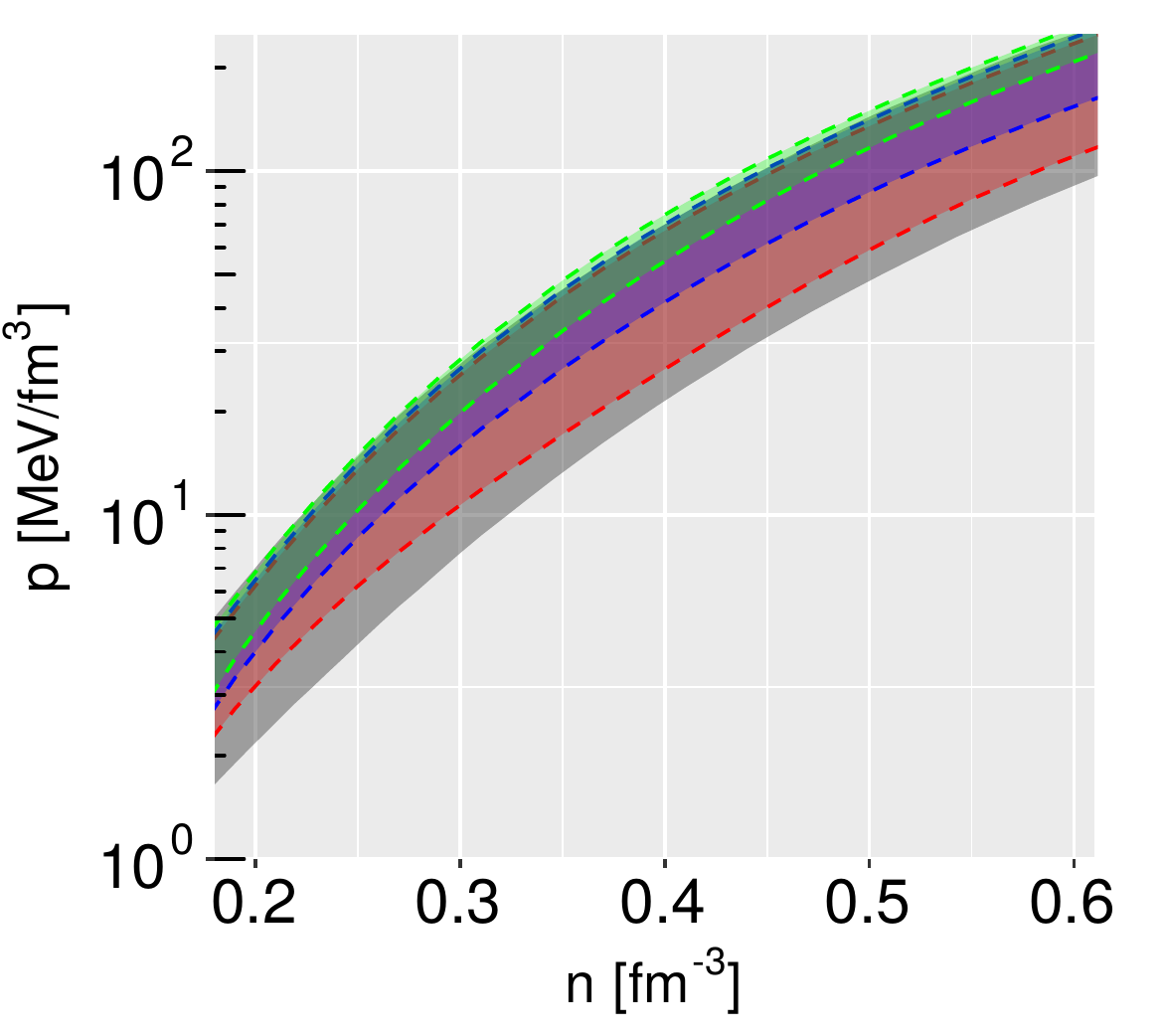}
	\includegraphics[width=0.9\columnwidth]{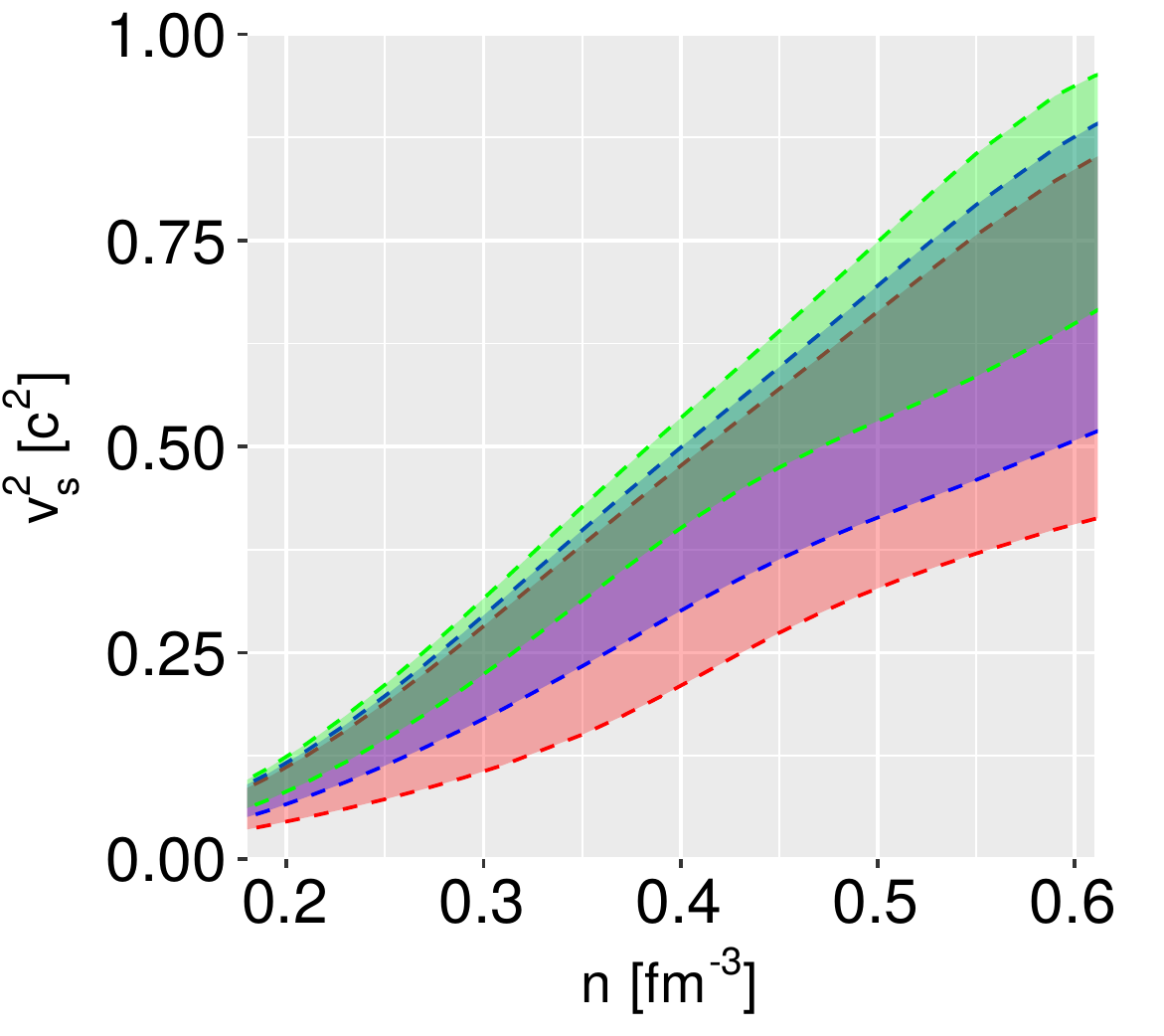}
	\caption{Inference results for the pressure (left) and speed of sound squared (right) as a function of baryonic density. The red, blue, and green regions indicate the  90\%  credible intervals for $M_{\text{max}}/M_{\odot}\in [2.0,2.66]$, $\in [2.2,2.66]$, and $\in [2.5,2.66]$, respectively. 
    The gray region corresponds to the 90\%  posterior credible level from \cite{Abbott18}.}
\label{fig:inference2}
\end{figure*}

Instead of a fixed value for $M_{\text{max}}$, we can consider an interval by determining $P(p(n))=\int_{a}^b P(p(n),M_{\text{max}}) dM_{\text{max}}$. The results are in Fig. \ref{fig:inference2}. Two main conclusions can be drawn: i) the detection of NS with a mass above two solar masses that decreases the NS maximum mass uncertainty interval from below will help constraining the high density EOS. Future observation from Square Kilometer Array (SKA) will certainly bring information on this lower limit; ii) taking as hypothesis a one branch mass-radius NS curve and imposing maximum masses above 2 solar masses does not allow to cover the whole pressure-density range defined by the LIGO/Virgo collaboration for the GW170817 binary NS merger.

\section{Conclusions}
\label{sec:conclusions}

In the present study we have  studied the possibility of constraining the EOS from the knowledge of the NS maximum mass, $M_{\text{max}}$, starting from the hypothesis of a EOS  with no first order phase transition and, as a second objective, we have shown how  to use the entire information obtained from the GW170817 event for the probability distribution of $\tilde{\Lambda}$ to make a probabilistic inference of the EOS.
We have generated a set of 150000 EOS that are constrained at low densities, saturation density and below, by nuclear properties. Otherwise, these EOS were required to be thermodynamically consistent, casual and to describe NS with maximum masses above 1.97 $M_\odot$. In a first step the EOS set was divided into three subsets corresponding to three intervals for the maximum mass, e.g. [1.97, 2.2]$M_\odot$, [2.2,2.4]$M_\odot$ and above 2.4$M_\odot$ and the pressure of catalysed beta-equilibrium neutral matter was obtained as a function of the baryonic density. It was shown that, while at low densities  the pressure band obtained for each set coincide below two times saturation density, above these density the differences are large and the pressure and speed of sound of the extreme sets (the lowest and largest mass sets) do not overlap, showing that the knowledge of the NS maximum mass may give some information on the high density EOS.  These sets, however, do not cover the whole EOS band predicted by the GW170817 event \cite{Abbott:2018exr}. The set we are using is more restrictive at low densities, since the EOS set used in the GW170817 analysis does not take into account nuclear properties. Recently some  tension between observational data and the nuclear physics inputs, and on the deformability probability distribution depending on the the inclusion or not of multimessenger information was discussed in \cite{Guven2020}. In particular, this analysis discusses the implications on the nuclear matter EOS properties.

We have next used a probabilistic approach based on our complete EOS dataset in order to use all the information the GW170817 gives us. This has allowed us to explore a region of the EOS phase space that was not accessible if nuclear matter properties close to saturation density are imposed. It  is shown that it is possible to extract constrained information on the high density density EOS from the knowledge of the maximum mass.
In particular, it was shown that the 90\% credible intervals of the pressure  as a function of the baryonic density obtained considering a maximum mass of  $2\,M_\odot$, $2.3\,M_\odot$ and $2.6\,M_\odot$ do not overlap for densities above 0.45 fm$^{-3}$. Moreover, neither the pressure nor the speed of sound 90\% credible intervals considering the two maximum masses,  $2\,M_\odot$ and $2.6\,M_\odot$ overlap for any density above saturation density. This seems to indicate that determining the NS maximum mass will give us strong constraints on the EOS at high densities. In particular,  the detection of more massive NS will narrow the uncertainty on the high density EOS.  It was also shown that the speed of sound increases monotonically well above the conformal limit, and that a maximum mass of  the order of 2.6$M_\odot$ may push the upper limit of the speed of sound to $\sim 1$ at 4 $n_0$. However, there is also a finite probability of obtaining an EOS that satisfies $v_s^2\lesssim 1/3 $ if the maximum mass is close to 2 $M_\odot$.\\

\section{Acknowledgments}

This work was partially supported by national funds from FCT (Fundação para a Ciência e a Tecnologia, I.P, Portugal) under the Projects No. UID/\-FIS/\-04564/\-2019, No. UID/\-04564/\-2020, and No. POCI-01-0145-FEDER-029912 with financial support from Science, Technology and Innovation, 
in its FEDER component, and by the FCT/MCTES budget through national funds (OE).

%\bibliography{biblio.bib}

%

\end{document}